\documentclass[prd,aps,a4paper,floatfix,amsmath,amssymb,twocolumn]{revtex4-1}
\usepackage[margin=3cm]{geometry}
\usepackage{graphicx,xcolor,array,float,fullpage}

\begin{document}

%%%%%%%%%%%%%%%%%%%%%%%%%%%%%%%%%%%%%%%%%%%%%%%%%%%%%%%%%%%%%

\title{Rigidly rotating perfect fluid stars in $2+1$ dimensions}

\author{Carsten Gundlach and Patrick Bourg}
\affiliation{Mathematical Sciences, University of Southampton,
  Southampton SO17 1BJ, United Kingdom}
\date{23 July 2020, revised 23 September 2020}

\begin{abstract}
Cataldo has found all rigidly rotating self-gravitating perfect fluid
solutions in 2+1 dimensions with a negative cosmological constant
$\Lambda$, for a density that is specified a priori as a function of a
certain radial coordinate. We rewrite these solutions in standard
polar-radial coordinates, for an arbitrary barotropic equation of
state $p(\rho)$. For any given equation of state, we find the
two-parameter family of solutions with a regular centre and finite
total mass $M$ and angular momentum $J$ (rigidly rotating stars). For
analytic equations of state, the solution is analytic except at the
surface, but including at the centre. Defining the dimensionless spin
$\tilde J:=\sqrt{-\Lambda}\,J$, there is precisely one solution for
each $(\tilde J,M)$ in the region $|\tilde J|-1<M<|\tilde J|$, which
consists of parts of the point particle region $M<-|\tilde J|$ and
overspinning regions $|\tilde J|>|M|$. In an adjacent compact part of
the black hole region $|\tilde J|<M$ (whose extent depends on the
equation of state), there are precisely two solutions for each
$(\tilde J,M)$. Hence exterior solutions exist in all three classes of
BTZ solution (black hole, point particle and overspinning), but not
all possible values of $(\tilde J,M)$ can be realised as
stars. Regardless of the values of $\tilde J$ and $M$, the causal
structure of all stars for all equations of state is that of
anti-de~Sitter space, without horizons or closed timelike curves.
\end{abstract}

\maketitle

\tableofcontents

%%%%%%%%%%%%%%%%%%%%%%%%%%%%%%%%%%%%%%%%%%%%%%%%%%%%%%%%%%%%%%%%%%%%%%%%%%

\section{Introduction}

%%%%%%%%%%%%%%%%%%%%%%%%%%%%%%%%%%%%%%%%%%%%%%%%%%%%%%%%%%%%%%%%%%%%%%%%%%

Classical Einstein gravity in 2+1 spacetime dimensions may appear to be
dynamically trivial because in 2+1 dimensions the Weyl tensor is
identically zero. This means that the full Riemann tensor is
determined by the Ricci tensor, and so by the stress-energy tensor of
the matter. Hence there are no gravitational waves, and the vacuum
solution is locally unique: Minkowski in the absence of a cosmological
constant $\Lambda$, de~Sitter for $\Lambda>0$, and anti-de~Sitter for
$\Lambda<0$.

However, in 1992, Ba\~nados, Teitelboim and Zanelli \cite{BTZ92} (from
now on, BTZ) noticed that 2+1 dimensional vacuum Einstein gravity with
$\Lambda<0$ admits rotating black hole solutions that are in close
analogy with the family of Kerr solutions in 3+1 dimensions. They can
be found easily by solving an axistationary ansatz for the metric, but
their existence was unexpected because the metric has to be locally
that of the 2+1-dimensional anti-de~Sitter solution (from now on,
adS3). In fact, these metrics can be derived as highly non-trivial
identifications of adS3 under an isometry \cite{BHTZ93}.

We define the cosmological length scale
\begin{equation}
\ell:=(-\Lambda)^{-{1\over 2}}.
\end{equation}
and the dimensionless spin
\begin{equation}
\tilde J:={J\over \ell}.
\end{equation}
The gravitational mass $M$ is already dimensionless in 2+1 dimensions. A key
difference to axistationary vacuum solutions in 3+1 dimensions is the
existence of a mass gap: while adS3 is given by the BTZ solution with
parameters $M=-1$ and $\tilde J=0$, only the BTZ solutions with $M>0$
and $|\tilde J|<M$ represent black holes.  Solutions with $-1<M<0$ and
$|\tilde J|<-M$ represent point particles, similar to those for
$\Lambda=0$ described in \cite{DeserJackiwTHooft84}. The status of
those with $|\tilde J|>|M|$, which we call ``overspinning'', remains
unclear.

The relevance of the BTZ solutions goes beyond vacuum because, roughly
speaking, the vacuum exterior of any rotating isolated object must be a
BTZ solution, even if the object itself is neither stationary nor
axisymmetric. 

More precisely, consider a region of spacetime with a timelike world
tube removed. We can make this region simply connected by making a cut
from the world tube to the outer boundary of the region. In the
resulting simply connected region the spacetime must be adS3. However,
when we make the region multiply connected again by identifying the
two sides of the cut, this identification is parameterised by an
isometry of adS3. The isometry group of adS3 is six-dimensional, but
it was shown in \cite{BHTZ93} that the gauge-invariant part of the
identification is characterised by only two parameters $(\tilde J,M)$,
parameterising precisely the BTZ solutions. A region of spacetime with
several world tubes removed requires one identification around each
world tube, and so is described by a pair $(\tilde J_i,M_i)$ for each
world tube representing a compact object.

By contrast, in 3+1 dimensions, the exterior of a rotating object is
not in general the Kerr solution, even if the object is axisymmetric
and stationary. The argument we have just given does not apply because
in more than 2+1 dimensions a vacuum spacetime need not be Minkowski
even locally. Put more physically, compact objects in 3+1 dimensions
can make not only their mass and spin, but also their internal
structure felt in their vacuum exteriors through tidal forces and
gravitational waves.

Perhaps the simplest example of axistationary matter solutions are
rotating perfect fluid stars. In this paper we examine if rigidly
rotating perfect fluid stars exist in 2+1 dimensions for reasonable
equations of state. Here we define a star to be a perfect fluid
solution with a regular centre and finite mass and spin. We allow both
for stars which have a surface at finite radius and are surrounded by
vacuum, and stars which fill all of space but whose density falls off
sufficiently rapidly. Given the existence of three different classes
of BTZ solutions, we ask if point-particle, black hole and
overspinning BTZ solutions can all be realised as exterior or
asymptotic spacetimes of rigidly rotating perfect fluid stars.

Hence in this paper we solve the Einstein-fluid equations
\begin{equation}
G_{ab} + \Lambda g_{ab}=8\pi T_{ab}
\end{equation}
with $\Lambda\le 0$ and the perfect-fluid stress-energy tensor
\begin{equation}
T_{ab}=(\rho+p)u_{a}u_{b}+pg_{ab},
\end{equation}
making an ansatz of stationarity and axisymmetry. The vector field
$u^a$ is tangential to the fluid worldlines, with $u^au_a=-1$, and $p$
and $\rho$ are the pressure and total energy density measured in the
fluid rest frame. We formally assume a barotropic equation of state
$p=p(\rho)$ given a priori. However, as we consider only axistationary
solutions, where all variables depend only on the radial coordinate
$r$, any solution with a given barotropic equation of state could also
a posteriori be a interpreted as a solution of a 2-parameter equation
of state $p=p(\rho,s)$ (where $s$ is, for example, the specific
entropy), together with a given stratification $s=s(\rho)$. We set
$c=G=1$ throughout.

Cruz and Zanelli \cite{Cruz95} have shown that static perfect fluid
solution require a non-positive cosmological constant $\Lambda \le 0$
and also studied in more detail the case of constant
energy-density. In \cite{Sa99}, the special cases of a polytropic
equation of state with and without cosmological constant were also
studied in \cite{Sa99} and \cite{Cornish91}. In \cite{Garcia03},
Garc\'ia et al. have derived all static circularly symmetric
spacetimes with $\Lambda \le 0$. Rigidly rotating configurations were
also studied \cite{Cornish94} and \cite{Lubo99}. Cataldo
\cite{Cataldo} has found all axistationary rigidly rotating perfect
fluid solutions in 2+1 spacetime dimensions with $\Lambda<0$. The
total energy density $\rho$ is specified a priori as a function of the
radial coordinate $\rho({\bar r})$. The metric and $p({\bar r})$ are
then given explicitly in terms of $\rho({\bar r})$ and four parameters
$C$, $D$, $E$ and $\omega_0$. The equation of state $p(\rho)$ is
implied only a posteriori by comparing $p({\bar r})$ and $\rho({\bar
  r})$. We summarise these results in Sec.~\ref{section:Cataldo}
below, followed by a list of questions that remained open: How does
one find the general solution if not $\rho({\bar r})$ but the equation
of state $p(\rho)$ is given a priori?  Which solutions have a regular
centre?  Which solutions have a vacuum exterior solution, and what is
its form?  What are the BTZ mass and angular momentum of such
star-like solutions?

To answer these questions, we translate Cataldo's solution into the
standard 2+1 form in terms of a lapse, shift and 2-metric, introduce
an area radius coordinate, identify Cataldo's radial coordinate
${\bar r}$ with a certain integral over the equation of state, and
identify the subset of solutions with a regular centre, which as
expected have only two free parameters (not four). We give expressions
for $M$ and $\tilde J$ in terms of these two parameters and certain
integrals involving only the equation of state.

Our solutions for a general equation of state are in implicit
form. They can be made explicit by evaluating an integral, inverting
the resulting function, and evaluating another integral. As already
obtained by Cataldo, this can be done for the linear equation of state
$p=\kappa\rho$ and the ``polytropic'' equation of state
$p=K\rho^k$. As a further example, we also consider the equation of
state $p=\kappa(\rho-\rho_s)$ for $\rho_s>0$.

%%%%%%%%%%%%%%%%%%%%%%%%%%%%%%%%%%%%%%%%%%%%%%%%%%%%%%%%%%%%%%%%%%%%%%%%%%

\section{General equation of state}

%%%%%%%%%%%%%%%%%%%%%%%%%%%%%%%%%%%%%%%%%%%%%%%%%%%%%%%%%%%%%%%%%%%%%%%%%%

\subsection{Rigidly rotating axistationary perfect fluid solutions}
\label{section:Cataldo}

%%%%%%%%%%%%%%%%%%%%%%%%%%%%%%%%%%%%%%%%%%%%%%%%%%%%%%%%%%%%%%%%%%%%%%%%%%

Cataldo \cite{Cataldo} has found axisymmetric, stationary, rigidly
rotating perfect fluid solutions of the Einstein equations in comoving
coordinates, defined by $u^a\propto(\partial_t)^a$, for a certain
choice of radial coordinate, in the form
\begin{equation}
ds^2=-\left({\bar r}\,d{\bar t}+\omega d\theta\right)^2 +h^{-1}\,d{\bar r}^2+h\,d\theta^2,
\end{equation}
where 
\begin{eqnarray}
\omega({\bar r})&:=&{\omega_0\over {\bar r}}+E{\bar r}, \\
\label{hdef}
h({\bar r})&:=&C-\Lambda {\bar r}^2+D {\bar r}+{\omega_0^2\over {\bar r}^2}
+16\pi \bar f({\bar r}), \\
\label{barfdef}
\bar f({\bar r})&:=& \int_{{\bar r}_0}^{\bar r} {\bar r}'\rho({\bar r}')\,d{\bar r}'
-{\bar r} \int_{{\bar r}_0}^{\bar r}\rho({\bar r}')\,d{\bar r}', \\
\label{pofr}
p({\bar r})&:=&{D\over 16\pi {\bar r}}-{1\over {\bar r}}\int_{{\bar r}_0}^{\bar r}\rho({\bar r}')\,d{\bar r}'.
\end{eqnarray}
Here ${\bar r}_0$ is an arbitray integration limit. (In solutions with
a regular centre, we will later choose it to correspond to the centre.)
These solutions are parameterised by the function $\rho({\bar r})$ and
the constants $\omega_0$, $E$, $C$ and $D$. (We denote the time and
radial coordinates of \cite{Cataldo} by ${\bar t}$ and ${\bar r}$ to
distinguish them from rescaled coordinates $t$ and $r$ that we
introduce below, and the area radius, which we will denote by $R$.)

At this point, it appears that the density $\rho$ has to be specified
as a function of the radial coordinate ${\bar r}$, which only
afterwards implies an equation of state $p(\rho)$ through the
expression (\ref{pofr}) for $p({\bar r})$. This issue was partly
addressed in \cite{Cataldo} by deriving explicit solutions for two
simple barotropic equations of state, but it remained unclear if and
how solutions can be obtained for an arbitrary equation of state
$p(\rho)$ given a priori.

It also remained unclear which solutions have a regular centre. This
issue was partly addressed in \cite{Cataldo} by giving explicit
solutions with a regular centre for the above-mentioned equations of
state. There was, however, no systematic construction of all solutions
with a regular centre for an arbitrary given equation of state in
terms of precisely two free parameters that control the mass and spin
of the star. Also lacking was a criterion on the equation of state for
a solution with a regular centre to either have a vacuum exterior, or
to be asymptotically adS3 with finite BTZ mass $M$ and spin $J$.

In the remainder of this paper, we resolve all these questions.

%%%%%%%%%%%%%%%%%%%%%%%%%%%%%%%%%%%%%%%%%%%%%%%%%%%%%%%%%%%%%%%%%%%%%%%%%%

\subsection{The equation of state}

%%%%%%%%%%%%%%%%%%%%%%%%%%%%%%%%%%%%%%%%%%%%%%%%%%%%%%%%%%%%%%%%%%%%%%%%%%

We first clarify the role of the equation of state. Differentiating
(\ref{pofr}), we obtain
\begin{equation}
\label{pODE}
{\bar r} {dp\over d{\bar r}}+p+\rho=0.
\end{equation}
Solving this separable ODE by integration, we find
\begin{equation}
\label{rceos}
\ln{{\bar r}\over{\bar r}_0}=-\int_{p_0}^{p({\bar r})}{dp\over p+\rho(p)}
=-\int_{\rho_0}^{\rho({\bar r})}{p'(\rho)\,d\rho\over p(\rho)+\rho},
\end{equation}
where $\rho_0:=\rho({\bar r}_0)$ is the density at ${\bar r}_0$ and
$p_0:=p(\rho_0)$ the corresponding pressure, given by the equation of
state $p(\rho)$. For stars, we will later choose ${\bar r}_0$ as the
value of ${\bar r}$ at the regular centre, so that $p_0$ is the
central pressure. 

Unless stated otherwise, we assume throughout that the equation of
state $p(\rho)$ is at least continuous and piecewise continuously
differentiable, with $0\le p'(\rho)<1$, and where $p'(\rho)=0$ is
allowed only at $p=0$. As a consequence, the sound speed
$\sqrt{p'(\rho)}$ is real and less than the speed of light, and the
inverse equation of state $\rho(p)$ also exists as a continuous
function that is piecewise once differentiable for $p>0$. We allow for
the possibility that $p(\rho_s)=0$ for some $\rho_s\ge 0$.

In obtaining (\ref{pODE}) by differentiating (\ref{pofr}) we have lost
the constant $D$. To find its value, we evaluate (\ref{pofr}) at
${\bar r}_0$, obtaining
\begin{equation}
\label{Dp0}
D=16\pi {\bar r}_0p_0.
\end{equation}

%%%%%%%%%%%%%%%%%%%%%%%%%%%%%%%%%%%%%%%%%%%%%%%%%%%%%%%%%%%%%%%%%%%%%%%%%%

\subsection{Standard form of the metric}

%%%%%%%%%%%%%%%%%%%%%%%%%%%%%%%%%%%%%%%%%%%%%%%%%%%%%%%%%%%%%%%%%%%%%%%%%%

For further analysis, we rearrange the metric in the usual 2+1 form,
and with the 2-metric expressed in terms of an area radius $R$, that
is, as
\begin{equation}
\label{polarradial}
ds^2=-\bar\alpha^2\,d{\bar t}^2
+a^2\left({dR\over d{\bar r}}\right)^2\,d{\bar r}^2
+R^2\,(d\theta+{\bar\beta}\,d{\bar t})^2,
\end{equation}
where $a$, $\bar\alpha$, $\bar\beta$ and $R$ are all functions of
${\bar r}$. Hence $\bar\alpha$ is the lapse, $\bar\beta$ the shift in
the angular direction, both with respect to the time coordinate ${\bar
  t}$, $g_{\theta\theta}=R^2$ defines the area radius $R$ as the
length of the Killing vector $\partial_\theta$ (and hence $R$ is a
scalar), and $g_{RR}=a^2$ if we use $R$ as the radial coordinate. We
read off
\begin{eqnarray}
R^2&=&h-\omega^2,\\
{\bar\beta}&=& -{{\bar r}\omega\over R^2}, \\
\bar\alpha^2&=&{\bar r}^2+R^2{\bar\beta}^2, \\
a^2&=&{1\over \left({dR\over d{\bar r}}\right)^2\,h}={4R^2\over \left({dR^2\over d{\bar r}}\right)^2\,h}
\end{eqnarray}
as functions of ${\bar r}$.  We see that ${\bar t}$ and ${\bar r}$
have nonstandard dimensions, namely length${}^{-1}$ and length${}^2$,
respectively. We use ${\bar r}_0$ to define a length scale
\begin{equation}
s:=\sqrt{{\bar r}_0}, 
\end{equation}
and then define
\begin{equation}
t:=s^2\,{\bar t} , \qquad r:={{\bar r}\over s}, 
\end{equation}
which have the usual dimension length. We correspondingly rescale the
lapse and shift as
\begin{equation}
\alpha:={\bar\alpha\over s^2}, \qquad \beta:={{\bar\beta}\over s^2}.
\end{equation}
The metric now takes the form
\begin{equation}
\label{trmetric}
ds^2=-\alpha^2\,dt^2+a^2\left({dR\over dr}\right)^2\,dr^2
+R^2\,(d\theta+\beta\,dt)^2.
\end{equation}

We introduce the dimensionless cosmological constant and spin
parameters
\begin{eqnarray}
\label{lambdadef}
\lambda&:=&s\sqrt{-\Lambda}\ge 0, \\
\Omega&:=&{\omega_0\over s^3},
\end{eqnarray}
and their combination
\begin{equation}
\label{mudef}
\mu:=\lambda^2-\Omega^2. 
\end{equation}
Note that $\lambda\ll1 $ corresponds to the length scale $s$ being
small compared to the cosmological length scale $\ell$,
but also, equivalently, to the cumulative effects of the cosmological
constant being small over length scales of size $s$. We will in
general consider $\lambda>0$, but at one point also $\lambda=0$,
interpreted as $\Lambda=0$. Otherwise, we always express
$\lambda$ in terms of the two independent parameters $\mu$ and $\Omega$.

To write all our equations in fully non-dimensional form, we
introduce the dimensionless radial coordinate $y$ and dimensionless
area radius $x$ defined by
\begin{equation}
y:={r\over s},\qquad x:={R\over s}.
\end{equation}
For a given equation of state $p(\rho)$ and reference density
$\rho_0$, the relation between the density $\rho$ and the
dimensionless radial coordinate $y$ is
\begin{equation}
\label{yrho0rho}
y(\rho_0;\rho)=\exp\left(-\int_{\rho_0}^{\rho}{p'(\tilde\rho)\,d\tilde\rho\over
  p(\tilde\rho)+\tilde\rho}\right),
\end{equation}
or equivalently 
\begin{equation}
\label{yp0p}
y(p_0;p)=\exp\left(-\int_{p_0}^{p}{d\tilde p\over
  \tilde p+\rho(\tilde p)}\right),
\end{equation}
where $\rho_0$ and $p_0=p(\rho_0)$ are the density and pressure at
$y=1$, $p'(\rho):=dp/d\rho$, and $\rho(p)$ is the inverse equation of
state, compare also Eq.~(50) of \cite{Lubo99}. We define the
  dimensionless function $f(y):=s^{-2}\bar f({\bar r})$, that is
\begin{equation}
\label{fdef}
f(y)=s^2\left( \int_1^y \rho(\tilde y)\tilde y\,d\tilde y
-y \int_1^y\rho(\tilde y)\,d\tilde y\right).
\end{equation}

We primarily use $s$ rather than $\ell$ to adimensionalise all other
variables and parameters in order to keep the limit $\Lambda=0$
regular. However, when we want to compare different solutions with
the same $\Lambda<0$, it is more natural to express the
dimensionful quantities $R$, $\rho$ and $p$ in terms of $\ell$, using
\begin{equation}
\label{sell}
s=\lambda\ell=\sqrt{\mu+\Omega^2}\,\ell.
\end{equation}
In particular we have
\begin{equation}
\label{RxmuOmega}
R=\ell\sqrt{\mu+\Omega^2}\,x
\end{equation}
and 
\begin{equation}
s^2\rho=(\mu+\Omega^2)\ell^2\rho.
\end{equation}

%%%%%%%%%%%%%%%%%%%%%%%%%%%%%%%%%%%%%%%%%%%%%%%%%%%%%%%%%%%%%%%%%%%%%%%%%%

\subsection{Local mass and angular momentum}

%%%%%%%%%%%%%%%%%%%%%%%%%%%%%%%%%%%%%%%%%%%%%%%%%%%%%%%%%%%%%%%%%%%%%%%%%%

For an arbitrary time-dependent axisymmetric spacetime in 2+1
spacetime dimensions, regardless of matter content, there exist two
conserved currents $\nabla_aj^a_{(J)}=0$ and $\nabla_aj^a_{(M)}=0$:
the conserved current due to the angular Killing vector, and a second,
more mysterious, one that generalises the Misner-Sharp mass that
exists for spherical symmetry in any dimension, to a conserved mass
that exists for axisymmetry in 2+1 dimensions only. In terms of the
metric (\ref{trmetric}), the corresponding conserved quantities are
given by
\begin{eqnarray}
\label{Jdefcoords}
J&=&{R^3{\partial\beta\over\partial r}\over {dR\over dr}a\alpha} , \\
\label{Mdefcoords}
M&=&{R^2\over\ell^2}+{J^2\over 4 R^2}-{1\over a^2}.
\end{eqnarray}
Note that these expressions hold in the axsymmetric but time-dependent
case. In the axistationary case that we consider here,
$\partial\beta/\partial r$ simply becomes $d\beta/dr$.  In any vacuum
region, $M$ and $J$ are constant with values equal to the BTZ
parameters of the same name, that is, the Einstein equations give
$M_{,r}=M_{,t}=J_{,r}=J_{,t}=0$. In particular, for constant $(J,M)$,
the polar-radial metric (\ref{trmetric}) takes the form
\begin{eqnarray}
\label{alphaBTZ}
c_0^2\alpha^2&=&-M+{R^2\over\ell^2}+{J^2\over 4R^2}, \\
a^2 &=&{1\over c_0^2\alpha^2}, \\
\label{betaBTZ}
c_0\beta&=&-{J\over 2R^2}+\beta_0.
\end{eqnarray}
We can further set $c_0=1$ by rescaling $t$ by the constant
factor $c_0$, and $\beta_0=0$ by a rigid rotation of the coordinate
system that corresponds to shifting $\theta$ by $\beta_0 t$. The
result is the standard form of the BTZ metric first given in
\cite{BTZ92}.

%%%%%%%%%%%%%%%%%%%%%%%%%%%%%%%%%%%%%%%%%%%%%%%%%%%%%%%%%%%%%%%%%%%%%%%%%%

\subsection{Solutions with a regular centre}

%%%%%%%%%%%%%%%%%%%%%%%%%%%%%%%%%%%%%%%%%%%%%%%%%%%%%%%%%%%%%%%%%%%%%%%%%%

We now demand that the solution has a regular centre at some value of
the radial coordinate ${\bar r}$. Without loss of generality we choose
the centre to be at the reference radius ${\bar r}_0$, so that
$R({\bar r}_0)=0$ and $\bar f({\bar r}_0)=(d\bar f/d{\bar r})({\bar
  r}_0)=0$. With these conditions, (\ref{hdef}) can be solved
  for the parameter $C$, which is now replaced as a free parameter by
  ${\bar r}_0$. 

We also demand that there is no conical singularity at the centre,
$a({\bar r}_0)=1$. However, a necessary condition for this limit to be
finite, given that $R({\bar r}_0)=0$ (by definition) and $(dR^2/d{\bar
  r})({\bar r}_0)\ne 0$ (by observation) is that $h({\bar r}_0)=0$,
and hence that $\omega({\bar r}_0)=0$. This last condition can be
solved for the parameter $E$. Applying l'H\^opital's rule, we then
have
\begin{equation}
\lim_{{\bar r}\to {\bar r}_0}a=\lim_{{\bar r}\to {\bar r}_0}{4\over
  {dR^2\over d{\bar r}}{dh\over d{\bar r}}}={4\over {dh\over d{\bar
        r}}({\bar r}_0)^2},
\end{equation}
and so we need $(dh/d{\bar r})({\bar r}_0)=2$, which can be solved for
$D$. The result, expressed for brevity in terms of our dimensionless
parameters $\mu$ and $\Omega$ and reference scale $s$, is
\begin{eqnarray}
\label{Erule}
E&=&-{\Omega\over s}, \\
\label{Crule}
C&=&s^2(\mu-2(1+\Omega^2)), \\
\label{Drule}
D&=&2(1-\mu).
\end{eqnarray}
For a given barotropic equation of state, the general solution with a
regular centre now has two dimensionless free parameters $\mu$,
$\Omega$, which govern, roughly speaking, the mass and spin of the
star. This is the number of free physical parameters one would expect
after imposing regularity at the centre. Note that, for fixed
$\Lambda$, $s$ is given in terms of $\mu$ and $\Omega$ by
(\ref{sell}), and from (\ref{Dp0}) and (\ref{Drule}), the central
  pressure is given in terms of $\mu$ by
\begin{equation}
\label{p0mus}
p_0={1-\mu\over 8\pi s^2},
\end{equation}
or equivalently
\begin{equation}
\label{p0muell}
p_0={1-\mu\over 8\pi (\mu+\Omega^2)\ell^2}. 
\end{equation}

The expression for the metric coefficients, for an arbitrary equation
of state, can be written concisely as
\begin{eqnarray}
\label{xyf}
x^2&=&\mu(y-1)^2+2(y-1)+16\pi f, \\
\label{alpha2xy}
\alpha^2&=&y^2+{\Omega^2(y^2-1)^2\over x^2}, \\
\label{a2xy}
a^2&=& {4y^2\over \left({dx^2\over dy}\right)^2\alpha^2}, \\
\label{betaxy}
\beta&=&{\Omega(y^2-1)\over sx^2}.
\end{eqnarray}
where $x$, $a$, $\alpha$ and $\beta$ are all functions of $y$. Note
  that $y\ge 1$ with $y=1$ at the regular centre. Recall
that $f(y)$ was defined in Eq.~(\ref{fdef}), where $\rho(y)$ is given
implicitly by inverting the integral (\ref{yrho0rho}), with the
integration limit $\rho_0=\rho(p_0)$ defined in terms of our free
parameters $\mu$ and $\Omega$ by Eq.~(\ref{p0muell}).

Eqns.~(\ref{yrho0rho}), (\ref{fdef}), (\ref{p0muell}) and
(\ref{xyf}-\ref{betaxy}) together fully specify our solutions, and can
be taken as the starting point for the analysis that follows.

For an analytic equation of state, $f(y)$ is analytic with
$f(y)=O(y-1)^2$ near the centre, and hence
\begin{eqnarray}
x^2&=&2(y-1)+O(y-1)^2, \\
\beta&=&{\Omega\over s}+O(y-1) \nonumber \\
&=&{\Omega\over\sqrt{\mu+\Omega^2}\,\ell}+O(y-1)
\end{eqnarray}
near the centre. We note for later use that, while $\beta$ is
proportional to $\Omega$ for small $\Omega$, it remains finite
everywhere as $|\Omega|\to\infty$.

We obtain a fully explicit solution in the radial coordinate $y$ if
and only if the integral (\ref{yrho0rho}) can be evaluated for
$y(\rho_0;\rho)$, this can then be inverted to give
$\rho(\rho_0;y)$, and if the integral (\ref{fdef}) can then also be
evaluated. Furthermore, we obtain a fully explicit solution in terms
of the area radius $R$ if and only if Eq.~(\ref{xyf}) can
also be inverted to give $y(x)$.

However, we do not need explicit solutions to establish analyticity of
the solution in the area radius $R$. In an open interval of $\rho$
where the equation of state $p(\rho)$ is analytic and $p+\rho>0$,
Eq.~(\ref{yrho0rho}) defines $y$ as a monotonically decreasing
analytic function of $\rho$ in this interval of $\rho$, and so
$\rho(y)$ exists and is analytic in the corresponding interval of
$y$. It follows that $f$ is an analytic function of $y$ in this
interval.  Hence $a$, $\alpha$ and $\beta$ are all analytic functions
of $y$ at least for $y>1$. A closer look shows that they are analytic
also at $y=1$, which corresponds to $x=0$. Moreover, $x^2$ is an
analytic function of $y$ for $y\ge 1$, and so implicitly $\rho$, $p$,
$a$, $\alpha$, $\beta$ are all analytic functions of $x^2$. In other
words, they are even analytic functions of $R$ for $R\ge 0$. For
typical equations of state, analyticity breaks down at the surface of
the star where $p=0$.

By a standard argument, analyticity in $R^2$ implies that if
we rewrite the metric in terms of Cartesian coordinates
$X:=R\cos\theta$, $Y:=R\sin\theta$, all coefficients of the metric in
the coordinates $(t,X,Y)$ are analytic functions of $X$ and $Y$ (and
independent of $t$), including at the centre $X=Y=0$.

The expressions for the local mass and angular momentum as functions
of $y$ are
\begin{eqnarray}
\label{My}
M&=&(\mu+2\Omega^2)x^2-{1\over 4}\left({dx^2\over dy}\right)^2
\nonumber \\ && -{\Omega^2(y^2-1){dx^2\over dy}\over y}, \\
\label{Jy}
J&=&s\Omega\left(2x^2-{(y^2-1){dx^2\over dy}\over y}\right),
\end{eqnarray}
or equivalently 
\begin{equation}
\tilde J=\sqrt{\mu+\Omega^2}\,\Omega
\left(2x^2-{(y^2-1){dx^2\over dy}\over y}\right).
\end{equation}
These are also even analytic functions of $R$.

%%%%%%%%%%%%%%%%%%%%%%%%%%%%%%%%%%%%%%%%%%%%%%%%%%%%%%%%%%%%%%%%%%%%%%%%%%

\subsection{The adS3 and test fluid cases}

%%%%%%%%%%%%%%%%%%%%%%%%%%%%%%%%%%%%%%%%%%%%%%%%%%%%%%%%%%%%%%%%%%%%%%%%%%

For $\mu=1$ the central pressure is zero, and so this must correspond
to the adS3 solution. Indeed, with $\mu=1$ the metric takes the form
\begin{eqnarray}
\label{xyvacuum}
x^2&=&y^2-1, \\
\label{alpha2vacuum}
\alpha^2&=&(1+\Omega^2)y^2-\Omega^2, \\
\label{a2vacuum}
a^2&=&\alpha^{-2}, \\
\label{beta0def}
\beta&=&s^{-1}\Omega=:\beta_0, 
\end{eqnarray}
and we have $M=-1$ and $J=0$. Hence this is the adS3 solution in a
rigidly rotating coordinate system, with constant angular velocity
$\beta_0$. In the vacuum solution, $\beta_0$ has no physical
significance, and can be set to zero.

Expanding in $\mu-1$, to leading order we obtain the test fluid limit,
in which a stationary, rigidly rotating, fluid configuration is held
together only by the cosmological constant (as well as being pulled
apart by rotation), but in which its self-gravity can be ignored. The
metric is that of adS3, but in a coordinate system that rotates with
the fluid. As in the self-gravitating case, the equation of state and
the central density $\rho_0$ implicitly determine a function
$\rho=\rho(\rho_0;y)$ through Eq.~(\ref{yrho0rho}). In the test fluid
case, from (\ref{xyvacuum}), $y$ is given in terms of the area radius
$R$, the cosmological constant $\Lambda$ and the constant angular
velocity $\beta_0$ as
\begin{equation}
y^2=1+x^2=1+R^2(-\Lambda-\beta_0^2),
\end{equation}
where we have used (\ref{lambdadef}), (\ref{mudef}) with $\mu=1$ and
(\ref{beta0def}) to eliminate $s$.  Hence we have an implicit
expression $\rho(R)$ for any rigidly rotating test fluid solution, for
arbitary central density $\rho_0$ and arbitrary constant angular
velocity $\beta_0$ (with respect to the Killing vector $\partial_t$),
given a cosmological constant $\Lambda<0$ and equation of state.

%%%%%%%%%%%%%%%%%%%%%%%%%%%%%%%%%%%%%%%%%%%%%%%%%%%%%%%%%%%%%%%%%%%%%%%%%%

\subsection{Star-like solutions}
\label{section:stars}

%%%%%%%%%%%%%%%%%%%%%%%%%%%%%%%%%%%%%%%%%%%%%%%%%%%%%%%%%%%%%%%%%%%%%%%%%%

We now look for solutions in which either $p=0$ occurs at finite
radius or $p\to 0$ and $\rho\to 0$ sufficiently rapidly as
$R\to\infty$ so that the solution has finite $M$ and $J$. We
  shall call such solutions ``stars''. Without any attempt at rigour,
we classify the possibilities by assuming that the fluid is
  polytropic at low pressure, that is
\begin{equation}
\label{surfaceapproximation}
p\sim \rho^k \quad\hbox{as}\quad p\to 0, 
\end{equation}
for some $k\ge 1$. We note that for $k<1$, the sound speed
$\sqrt{p'(\rho)}$ diverges as $\rho\to 0$. We therefore disregard this
range as unphysical.

From (\ref{p0muell}), we require $\mu\le 1$ for the central pressure
to be non-negative, and from (\ref{xyf}) we further require $\mu\ge 0$
for $x(y)$ to be a monotonically increasing function for all $y$, in
particular at large $y$. Stars therefore exist only with $\Lambda<0$,
and for $0\le \mu\le 1$. Physically, from (\ref{mudef}), $\mu>0$ means
that the Hubble acceleration is centripetal ($\Lambda<0$) and larger
than the centrifugal acceleration due to the rigid rotation
($\lambda^2>\Omega^2$). Both the Hubble and the centrifugal
acceleration depend on radius in the same way, and so this is true
either for all $y$ or for none.

\paragraph{Stars with a surface} From (\ref{yp0p}), we see that 
the solution has a
surface $p(y_*)=0$ at some finite coordinate radius $y_*$ and finite
area radius $x_*$ if and only if the integral
\begin{equation}
\label{ystareos}
y_*(p_0):=y(p_0;0)=\exp\int_0^{p_0}{dp\over p+\rho(p)}.
\end{equation}
converges. Note that in this case $y_*(0)=1$. In the approximation
(\ref{surfaceapproximation}) this is the case for $k>1$. The limiting
case $k=\infty$ can be interpreted as a fluid where $\rho=\rho_s>0$ is
finite at $p=0$. (One may think of such a perfect fluid as a liquid,
rather than a gas).

In the exterior $y>y_*$, the solution must be equal to a BTZ solution
with constant $M$ and $J$. To verify this, we note that in the
exterior, (\ref{fdef}) reduces to
\begin{equation}
\label{fext}
16\pi f=m-2(1-\mu)y,
\end{equation}
where we have defined the integrated fluid mass
\begin{equation}
\label{mdef}
m:=16\pi s^2\int_1^{y_*}\rho\,y\,dy.
\end{equation}
We have identified the coefficient of $y$ in (\ref{fext}) as
  $-D$ by demanding that (\ref{pofr}) holds in the vacuum region
$p=0$, and have then used (\ref{Drule}) to eliminate $D$. 

As $y\ge 1$ in the integral in (\ref{mdef}), we have 
\begin{equation}
\label{minequality}
m\ge 16\pi s^2\int_1^{y_*}\rho\,dy=2(1-\mu),
\end{equation}
where to obtain the last equality we have evaluated (\ref{pofr})
in the vacuum region $p=0$ and used (\ref{Drule}). 

To clarify what free parameters determine $m$, we use (\ref{yp0p}) to
eliminate $y$ and (\ref{p0mus}) to eliminate $s$ in favour of the
central pressure $p_0$, and then (\ref{p0muell}) to in turn express
$p_0$ in terms our free parameters $\mu$ and $\Omega$. We obtain
\begin{equation}
m=2(1-\mu)\,I\left({1-\mu\over 8\pi (\mu+\Omega^2)\ell^2}\right),
\end{equation}
where
\begin{equation}
\label{Ip0def}
I(p_0):=\int_0^{p_0}\exp\left(-2\int_{p_0}^{p}{d\tilde p\over
  \tilde p+\rho(\tilde p)}\right){\rho(p)\over p_0}{dp\over
  p+\rho(p)}.
\end{equation}
So in general $m$ depends on $\mu$ and $\Omega^2$, as well as of
course on the equation of state. Note that from (\ref{minequality}),
we have $I(p_0)\ge 1$. 

To simplify the expressions that follow, we
define the auxiliary quantity
\begin{equation}
A(\mu,\Omega):=m(\mu,\Omega)+\mu-2.
\end{equation}
By definition, $A(1,\Omega)=-1$ in the vacuum or test fluid case, where
$m=0$. From (\ref{minequality}), we have 
\begin{equation}
\label{Ainequality}
A+\mu\ge 0.
\end{equation}

With $f$ given by (\ref{fext}), the metric coefficients in the vacuum
exterior are given by (\ref{xyf}-\ref{betaxy}) as
\begin{eqnarray}
\label{x2yexterior}
x^2&=&\mu y^2+A, \\
\alpha^2&=&y^2+{\Omega^2(y^2-1)^2\over \mu y^2+A}, \\
a^2&=& {1\over\mu^2\alpha^2},\\
\label{betayexterior}
\beta&=&{\Omega(y^2-1)\over s(\mu y^2+A)}.
\end{eqnarray}

Substituting (\ref{x2yexterior}) into the expressions (\ref{My}) and
(\ref{Jy}) for $M$ and $J$, we obtain the constant values
\begin{eqnarray}
\label{Mtot}
M&=&M_{\rm tot}:=A\mu+2(A+\mu)\Omega^2, \\
\label{Jtildetot}
\tilde J&=&\tilde J_{\rm tot}:=2\sqrt{\mu+\Omega^2}(A+\mu)\Omega,
\end{eqnarray}
or equivalently
\begin{equation}
\label{Jtot}
J_{\rm tot}=2s(A+\mu)\Omega.
\end{equation}
It is then easy to verify that the exterior metric
(\ref{x2yexterior}-\ref{betayexterior}), is
(\ref{alphaBTZ}-\ref{betaBTZ}), generally with $c_0\ne 1$ and
$\beta_0\ne 0$.

\paragraph{Stars without a surface} If the integral (\ref{ystareos})
diverges but the integral (\ref{mdef}) with $y_*=\infty$ converges to
a finite value of $m$, the star has no surface but finite mass.

Taking the limit of $M(y)$ and $J(y)$ as $y\to\infty$, we again obtain
the finite total values given by (\ref{Mtot}) and (\ref{Jtot}). The
metric is now asymptotic (rather than strictly equal) to the BTZ
metric (\ref{trmetric},\ref{alphaBTZ}-\ref{betaBTZ}).

In these stars without a sharp surface, we can nevertheless roughly
identify a central region where self-gravity of the star is important
and $M$ and $|J|$ still increase, and an outer region, or stellar
atmosphere, where $M$ and $J$ are essentially constant and the fluid
is essentially a test fluid on the BTZ spacetime with parameters
$M_{\rm tot}$ and $J_{\rm tot}$.

In our approximation (\ref{surfaceapproximation}) this happens in the
marginal case $k=1$, we need to also specify the constant of
proportionality, as the dimensionless parameter $\kappa$ in
\begin{equation}
p\simeq\kappa\rho \quad\hbox{as}\quad p\to 0,
\end{equation}
for some $0<\kappa<1$. The pressure and density fall off as $\rho\sim
p\sim y^{-1-{1\over \kappa}}$, and so once again $m$ is finite, but
there is now no surface at finite radius, and the metric is only
asymptotically BTZ, with $y_*=\infty$. The sound speed is also less
than the the speed of light for $0<\kappa<1$.

\paragraph{Non-stars} 
When not only $y_*$ but $m$ diverges, $f(y)$ grows faster than $y$ as
$y\to\infty$. In the approximation (\ref{surfaceapproximation}), this
is the case for $1/2\le k<1$, when $\rho\sim y^{-{1\over k}}$ and
$f\sim y^{2-(1/k)}$ as $y\to\infty$. However, we have already ruled
out $k<1$ on the grounds that the sound speed $\sqrt{p'(\rho)}$
diverges at the surface. The expressions for $M(y)$ and $J(y)$ also
diverge, and so the spacetime is not asymptotically BTZ. Such
solutions do not describe stars. Recall again that we have already
ruled out $k<1$ on the grounds of diverging sound speed.

%%%%%%%%%%%%%%%%%%%%%%%%%%%%%%%%%%%%%%%%%%%%%%%%%%%%%%%%%%%%%%%%%%%%%%%%%%

\subsection{The manifold of solutions}

%%%%%%%%%%%%%%%%%%%%%%%%%%%%%%%%%%%%%%%%%%%%%%%%%%%%%%%%%%%%%%%%%%%%%%%%%%

In contrast to 3+1 and higher dimensions, the vacuum exterior metric,
or the asymptotic metric at infinity, of a rotating star is given by a
BTZ metric. It is therefore of interest what region in the $(\tilde
J,M)$ plane is covered by possible stellar exterior solutions. Recall
that for stars the parameters $\mu$ and $\Omega$ can take any values
in the strip
\begin{equation}
0<\mu\le 1, \quad -\infty<\Omega<\infty.
\end{equation}
In the following, we suppress the suffix ``tot'' for brevity, and for
the rest of this Section, $M$ and $\tilde J$ always denote the total
mass and spin of the spacetime, measured at infinity.

The manifold of solutions is uniquely parameterised by
$(\Omega,\mu)$. However, if we are interested more in the values of
$(\tilde J,M)$, we can present the solution manifold as a hypersurface
in $(\tilde J,M,\mu)$ space. The case of the linear equation of state
$p=\kappa\rho$ is non-generic in that $A$ is a function of $\mu$ only,
but it, and in particular the value $\kappa=1/2$, can serve as a
concrete illustration of the general considerations presented
below. The solution manifold parameterised by $(\Omega,\mu)$ for the
equation of state $p=\rho/2$ is shown in Fig.~\ref{fig:fig1}. The same
solution manifold embedded in $(\tilde J,M,\mu)$ space is shown in
Fig.~\ref{fig:fig2}, and the projection of this embedding down into
the $(\tilde J,M)$ plane in Fig.~\ref{fig:fig3}. We stress that the
following arguments hold for all equations of state that admit
star-like solutions, and so these figures apply qualitatively to all
equations of state.

\paragraph{Boundary $\mu=1$ of solution space}

We have already seen that $\mu=1$ at finite $\Omega$ (the thick black
line in Fig.~\ref{fig:fig1}) corresponds to a rotating test fluid on
the adS3 spacetime with $M=-1$ and $\tilde J=0$. However, taking the
simultaneous limit $\mu\to 1_-$, $\Omega\to\pm\infty$ of (\ref{Mtot})
and (\ref{Jtildetot}) such that 
\begin{equation}
\mu=1-{\tilde q\over \Omega^2}
\end{equation}
for some fixed constant $\tilde q>0$, we have
$s\to\infty$ and $p_0\to 0$ and so, for finite $I(0)$, we obtain
\begin{equation}
m\simeq 2(1-\mu)\,I(0),
\end{equation}
giving
\begin{equation}
A+\mu\simeq {q\over \Omega^2}, \quad q:=2[I(0)-1]\tilde q,
\end{equation}
and hence two 1-parameter families of solutions with
\begin{equation}
\label{chevronlowerboundary}
M=-1+q, \quad \tilde J=\pm q. 
\end{equation}
 From (\ref{Ainequality}), we have that $q\ge 0$. See the blue region
 in Fig.~\ref{fig:fig1} as $\Omega\to\pm\infty$, and the thick dashed
 black line in Fig.~\ref{fig:fig2}. In this limit, the fluid is
 infinitely dilute but infinitely extended. Note that even though
 $\Omega\to\infty$, the angular velocity $\beta$ is finite
 everywhere. The integrated fluid rest mass $m$ vanishes, but
 $M>-1$. Intuitively, this nontrivial gravitational mass comes from
 rotational energy.

We now show, assuming an analytic equation of state for small $p>0$,
that $I(0)=1$ if the star has a surface at finite radius. To see this,
we write
\begin{equation}
I(p_0) = \int_0^{p_0} y^2(p_0;p) \frac{\rho}{p_0} \frac{dp}{p+\rho}\ge
0.
\end{equation}
We can bound $1\le y^2\le y_*^2$ in the integrand, and so
\begin{equation}
\frac{1}{p_0} \int_0^{p_0} \frac{\rho}{p+\rho} dp \le I(p_0) \le 
\frac{y_\star^2}{p_0} \int_0^{p_0} \frac{\rho}{p+\rho} dp.
\end{equation}
From $y_*(0)=1$ (as noted above) and the squeeze theorem, we then have
\begin{equation}
I(0) = \lim\limits_{p_0 \to 0} \frac{1}{p_0} \int_0^{p_0} {dp\over 
1+{p\over\rho}} \label{Ip0limitintegral}.
\end{equation}
From causality, $p/\rho$ must remain bounded as $p\to 0$. If in fact
$p/\rho\to 0$ as $p\to 0$, we have $I(0)=1$.

In the other case, where $p/\rho\to \kappa$ remains finite as $p\to
0$, the surface of the star is at infinity and so we cannot rely on
(\ref{Ip0limitintegral}). However, one can see by explicit calculation
that $I(0)=1/(1-\kappa)$ for this case, which is finite, see
also (\ref{mlineos}) below.

\paragraph{Boundary $\mu=0$ of solution space}

If $A(0,\Omega)$ is finite, the boundary $\mu=0$ of solution
  space corresponds to a family of solutions with
\begin{equation}
\label{chevronupperboundary}
M=2A(0,\Omega)\,\Omega^2, \quad \tilde J=2A(0,\Omega)\,|\Omega|\,\Omega,
\end{equation}
Note that $A(0,\Omega)\ge 0$ from (\ref{Ainequality}), and so these
solutions obey $M\ge 0$ with $|\tilde J|=M$. See the thick blue line
in Figs.~\ref{fig:fig1} and \ref{fig:fig2}.

\paragraph{Second family of critically spinning solutions}

There is a second family of solutions with $|\tilde J|=|M|$, over a
finite range of $M$ including both positive and negative values of
$M$, namely
\begin{equation}
\Omega=\pm\Omega_c(\mu), 
\label{greencurves}
\end{equation}
where $\Omega_c(\mu)$ is defined by solving
\begin{equation}
A^2=4(A+\mu)\Omega^2
\end{equation}
for $\Omega^2$, given $\mu$.
Along these curves, parameterised by $\mu$, we have
\begin{equation}
|M|=|\tilde J|=A\left(\mu+{A\over 2}\right).
\end{equation}
The range $1>\mu>0$ corresponds to the range $-1/2<M<M_0$. Here 
\begin{equation}
M_0:=8\Omega_0^2, 
\end{equation}
where $\Omega_0$ is the positive solution of 
\begin{equation}
A(0,\Omega_0)=4\Omega_0^2.
\end{equation}
[Note that therefore $\Omega_0=\Omega_c(0)$.] See the thick green
lines in Figs.~\ref{fig:fig1} and \ref{fig:fig2}.. The two
curves intersect at $M=\tilde J=0$, which corresponds to $\mu=\mu_c$
defined by 
\begin{equation}
\label{mucdef}
A(\mu_c,0)=0.
\end{equation}
This always has a solution in the range $0\le\mu_c<1$ because
$A(\mu,\Omega)$ is continuous with $A(0,\Omega)\ge 0$ and
$A(1,\Omega)=-1$. [We assume without proof that there is only one
  solution.] At their upper ends, the two curves are asymptotic to
$\mu=1$ as $\Omega\to\pm \infty$ in the $(\Omega,\mu)$ strip, but in
the $(\tilde J,M)$ plane they end at the finite points $M=-1/2$,
$\tilde J=\pm 1/2$. At their lower ends they intersect $\mu=0$ at
finite $|\Omega|=\Omega_0$, corresponding to $|\tilde{J}|=M=M_0>0$.

\paragraph{Double cover of a region in the $(\tilde J,M)$ plane}

As there are two solutions for $M=|\tilde J|$ for $0\le M<M_0$, by
continuity there must be a region of the $(\tilde J,M)$ plane that is
doubly covered by the manifold of solutions. As the solutions
$M=|\tilde J|$ corresponding to $\mu=0$ lie on one boundary of the
solution manifold, they also form one boundary of the doubly-covered
region [in $(\Omega,\mu)$ and $(\tilde J, M)$, respectively]. The
other boundary of the doubly-covered region in the $(\tilde J,M)$
plane occurs where the solution manifold of
  Fig.~\ref{fig:fig2} folds over. This occurs where
\begin{equation}
\left|{\partial(\tilde J,M)\over\partial(\Omega,\mu)}\right|=0,
\end{equation}
which is equivalent to 
\begin{equation}
2(A+\mu)(\mu A_{,\mu}+A-4\Omega^2)+(A-4\Omega^2-3\mu)\Omega
A_{,\Omega}=0.
\end{equation}
This implicitly defines a curve
\begin{equation}
\label{redcurve}
\Omega=\pm \Omega_r(\mu), \quad 0<\mu<\mu_r,
\end{equation}
where $\mu_r$ is defined by $\Omega_r(\mu_r)=0$, giving
\begin{equation}
\mu_rA_{,\mu}(\mu_r,0)+A(\mu_r,0)=0.
\end{equation}
[Note that $\Omega_r(0)=\Omega_0$. We assume without proof that there
  is only one such curve, that is, the solution manifold is not folded
  over more than double.]

In fluid parameter space $(\Omega,\mu)$, the doubly-covered region
lies between the curves (\ref{greencurves}) for $0<\mu<\mu_c$ (the
lower part of the two green curves in Fig.~\ref{fig:fig1}) and the
curve $\mu=0$ for $-\Omega_0<\Omega<\Omega_0$ (part of the blue
line). It is divided into two halves by (\ref{redcurve}) (the red
curve). All three curves intersect at the two points $\mu=0$,
$\Omega=\pm \Omega_0$. Pairs of points from those two halves of the
doubly-covered region have the same values of $M$ and $\tilde J$.

In BTZ parameter space $(\tilde J,M)$, the doubly covered region lies
between $|\tilde J|=M$ for $0<M<M_0$ (corresponding to both the
blue and green curves in Fig.~\ref{fig:fig2}), and the red curve
\begin{equation}
\label{tilderedcurve}
\Omega=\pm \tilde\Omega_r(M), \quad 0<M<M_0,
\end{equation}
which is given implicitly by (\ref{Mtot}) and (\ref{Jtildetot}) with
(\ref{redcurve}). The double cover becomes clearer by comparing
Fig.~\ref{fig:fig2} with its top view, Fig.~\ref{fig:fig3}.  The
corner points at $\mu=0$, $\Omega=\pm\Omega_0$ have $M=|\tilde
J|=M_0$. Hence the maximum possible $M$ for given $|\tilde J|<M_0$ is
obtained on the red curve. In particular, the maximum possible mass
without rotation is given by $\Omega=0$ and $\mu=\mu_r$, and is
\begin{equation}
M_r:=M(\mu_r,0)=A(\mu_r,0)\mu_r.
\end{equation}

The red curve (\ref{redcurve}) corresponds to a curve of solutions
that have a zero mode, a static linear perturbation that corresponds
to an infinitesimal change of $(\mu,\Omega)$ that leaves $(\tilde
J,M)$ invariant to linear order. This signals that a linear perturbation
mode changes from stable to unstable across the red curve. This is
familiar from nonrotating stars in 3+1 dimensions, where an extremum
of the mass as a function of central density signals a separation
between stable and unstable stars, with the less dense stars stable
and the more dense ones unstable. We conjecture that the solutions in
the doubly-covered region with smaller $\mu$ (and hence larger
  central density) are unstable, corresponding to region below the
red curve in Fig.~\ref{fig:fig1}. As their asymptotic metrics are of
black-hole type, it is possible that these unstable solutions collapse
to a black hole when perturbed in a suitable way.

We have obtained some evidence for this conjecture by time-evolving
the two solutions with the equation of state $p=\rho/2$ represented by
the orange and black dots in Fig.~\ref{fig:fig1}. Adding a small
perturbation of the density with either sign to the less dense
(orange) solution sets up propagating perturbations that remain
small. Adding a small density perturbation to the denser (black) solution
results in a highly nonlinear oscillation for one sign of the
perturbation, where the central density repeatedly decreases below
that of the orange solution, while perturbing the initial density with
the opposite sign triggers prompt collapse to a black hole.

\paragraph{Summary of $\Lambda<0$}

In summary, the manifold of solutions contains a unique
  solution with given $(\tilde J,M)$ in the chevron-shaped region
\begin{equation}
\label{chevron}
|\tilde J|-1<M<|\tilde J|
\end{equation}
that is bounded by the curves (\ref{chevronlowerboundary}) and
(\ref{chevronupperboundary}), while in a contiguous compact region
bounded by $|\tilde J|=M$ for $0<M<M_0$ and the curve (\ref{redcurve})
there are two solutions with the same given $(\tilde J,M)$. There
are no solutions with $(\tilde J,M)$ outside these two regions.

\paragraph{The case $\Lambda=0$}

We now consider the limit where the length scale $s$ remains finite
but $\Lambda\to 0$. Then $\lambda^2=\mu+\Omega^2=0$, so in this limit
$\mu=\Omega=0$. Therefore, no rigidly rotating stars can
exist. Intuitively, only the cosmological contraction due to
$\Lambda<0$ can balance the centrifugal acceleration of rigid
rotation, while the curvature generated by stress-energy cannot.
Setting $\Omega=0$, replacing $(\mu+\Omega^2)\ell^2$ with $s^2$, and
then setting $\mu=0$, we obtain
\begin{equation}
m=2I\left(1\over 8\pi s^2\right).
\end{equation}
Eqns.~(\ref{yrho0rho}) and (\ref{fdef}) still hold, and so do
(\ref{xyf}-\ref{betaxy}) and (\ref{My}), reduced to
\begin{eqnarray}
\label{xynoLambda}
x^2&=&2(y-1)+16\pi f, \\
\alpha^2&=&y^2, \\
\label{MxynoLambda}
M&=&-{1\over a^2}=-{1\over 4}\left({dx^2\over dy}\right)^2,
\end{eqnarray}
with $\beta=0$ and $J=0$.  They define an analytic interior solution
for analytic equation of state, with in particular a regular
centre. However, in the vacuum exterior to this interior
solution, (\ref{xynoLambda}) with (\ref{fext}) gives $x^2=m-2$, which
is constant, so from (\ref{MxynoLambda}) $M=0$. This means that $a$
diverges at the surface, but the metric expressed in terms of $y$
remains regular, and in the exterior it is
\begin{equation}
ds^2=-y^2\,dt^2+s^2\left({dy^2\over m-2}+(m-2)\,d\theta^2\right),
\end{equation}
for $y_*<y<\infty$. The spatial geometry is a cylinder, see also
Eq.~(79) of \cite{Lubo99}. If $y_\star$ is finite, we do not consider
such a solution as a star.

%%%%%%%%%%%%%%%%%%%%%%%%%%%%%%%%%%%%%%%%%%%%%%%%%%%%%%%%%%%%%%%%%%%%%%%%%%

\subsection{Causal structure}

%%%%%%%%%%%%%%%%%%%%%%%%%%%%%%%%%%%%%%%%%%%%%%%%%%%%%%%%%%%%%%%%%%%%%%%%%%

If we apply the standard compactification of adS3, namely
\begin{equation}
R=\ell\tan{\psi\over\ell},
\end{equation}
to the BTZ metric in its standard form, (\ref{trmetric},\ref{alphaBTZ}-\ref{betaBTZ})
with $c_0=1$ and $\beta_0=0$, we obtain
\begin{eqnarray}
ds^2&=&{1\over \cos^2{\psi\over\ell}}\Biggl[
-F\,dt^2 
+G^{-1}\,d\psi^2 \nonumber \\ &&
+\ell^2\sin^2{\psi\over\ell}
\left(d\theta+H\,dt\right)^2 \Biggr], 
\end{eqnarray}
where $F=G$ and
\begin{eqnarray}
G&=&1-(M+1)\cos^2{\psi\over\ell}
+{J^2\cos^4{\psi\over\ell}\over 4\sin^2{\psi\over\ell}}, \\
H&=&{J\cos^2{\psi\over\ell}\over 2\sin^2{\psi\over\ell}}.
\end{eqnarray}
This is conformal to a metric (the one in the large square brackets)
that is regular everywhere, or in the black-hole case everywhere
outside the event horizon, but always including at $\psi/\ell=\pi/2$,
which is therefore revealed as a timelike conformal boundary.  In our
star-like solutions, $F\ne G$ and $H$ are different functions from
those given above, but they are finite and non-zero for
$0\le\psi/\ell\le \pi/2$.

For the BTZ metrics corresponding to black holes, the familiar Penrose
diagram \cite{BHTZ93} is a different one, being a square that is
compact in the time as well as the radial direction. At first sight,
this seems to contradict the above conformal picture for a star, in
which the conformal metric has an infinite range of $t$. The apparent
contradiction is resolved by noticing that the black hole conformal
diagram contains at its top and right corner a point representing
timelike infinity where the curve representing the future branch of
the event horizon meets the curve representing the timelike conformal
boundary. If we now cover up the black hole region with a star, the
timelike curve representing the surface of the star and the timelike
conformal boundary meet at the same point in the conformal
diagram. Both have infinite proper length, and are tangential to the
stationary Killing vector. Moreover, a radial light ray reflected at
both curves travels between them an infinity number of times before
reaching the point in the conformal diagram where they meet. Hence
there must be a conformal transformation where these two curves remain
parallel and have infinite coordinate length in the resulting Penrose
diagram, as derived above.

A second question about the causal structure is if the spacetime
admits closed timelike curves. It is obvious that closed timelike
curves exist if there is a region where the metric coefficient
$g_{\theta\theta}=R^2$ is negative. Conversely, Ba\~nados, Henneaux,
Teitelboim and Zanelli \cite{BHTZ93} have proved that the BTZ metrics
do not contain closed timelike curves if there is no region with
$R^2<0$, or if such regions are excluded. The proof only relies on the
signature of the metric coefficients, not their form, and so
generalizes to metrics of the form (\ref{polarradial}), as long as
$a^2$ and $\alpha^2$ remain positive. Hence, as $a^2$, $\alpha^2$ and
$R^2$ are manifestly non-negative in our star-like solutions, they do
not contain closed timelike curves. (The examples of solutions with
closed timelike curves given by Cataldo \cite{Cataldo} can therefore
not be star-like, that is, have both a regular centre and be
asymptotically BTZ.)

%%%%%%%%%%%%%%%%%%%%%%%%%%%%%%%%%%%%%%%%%%%%%%%%%%%%%%%%%%%%%%%%%%%%%%%%%%

\section{Simple equations of state}

%%%%%%%%%%%%%%%%%%%%%%%%%%%%%%%%%%%%%%%%%%%%%%%%%%%%%%%%%%%%%%%%%%%%%%%%%%

%%%%%%%%%%%%%%%%%%%%%%%%%%%%%%%%%%%%%%%%%%%%%%%%%%%%%%%%%%%%%%%%%%%%%%%%%%

\subsection{Ultrarelativistic linear equation of state $p=\kappa\rho$}

%%%%%%%%%%%%%%%%%%%%%%%%%%%%%%%%%%%%%%%%%%%%%%%%%%%%%%%%%%%%%%%%%%%%%%%%%%

In the following, we concentrate on solutions with the
ultra-relativistic (linear) equation of state $p=\kappa\rho$, assuming
the physical range $0<\kappa<1$ of the equation of state parameter,
which gives a real speed of sound smaller than the speed of light.
(With the value $\kappa=1/2$ in particular this equation of state can
be interpreted as a gas of massless particles without internal degrees
of freedom.) We have already seen above that star-like solutions with
this equation of state have no surface at finite radius but are
asymptotically BTZ. From (\ref{yrho0rho}) we have
\begin{equation}
\label{rhoylineos}
\rho(\rho_0;y)=\rho_0y^{-{1+\kappa\over\kappa}},
\end{equation}
and hence from (\ref{fdef})
\begin{equation}
8\pi f(y)=(1-\mu)\left((1-y)
+{\kappa\over 1-\kappa}\left(1-y^{-{1-\kappa\over\kappa}}\right)\right).
\end{equation}
Of the metric coefficients, we here write out only 
\begin{equation}
\label{x2ylineos}
x^2=\mu(y^2-1)+{2\kappa(1-\mu)\over
  1-\kappa}\left(1-y^{-{1-\kappa\over\kappa}}\right).
\end{equation}
The other metric coefficients are given by
(\ref{alpha2xy}-\ref{betaxy}). 

In the test fluid case $\mu=1$ we have $x^2=y^2-1$, and so the density
in terms of the area radius takes the simple form
\begin{equation}
\rho=\rho_0\left[1+R^2(-\Lambda-\beta_0^2)\right]^{-{1+\kappa\over 2\kappa}},
\end{equation}
where the central density $\rho_0$ is arbitrary (but assumed so small
that self-gravity can be neglected) and $\beta_0$ is the constant
angular velocity.

Integrating (\ref{rhoylineos}), we have
\begin{equation}
\label{mlineos}
m=2{1-\mu\over1-\kappa} \quad \Leftrightarrow \quad
I(p_0)={1\over1-\kappa},
\end{equation}
and so
\begin{equation}
A={2\kappa-(1+\kappa)\mu\over 1-\kappa}.
\end{equation}
For this particular equation of state, $I(p_0)$ is constant, and
so $m$ and $A$ depend on $\mu$ only but (untypically) not on
$\Omega$. The total mass and spin at infinity are
\begin{eqnarray}
M_{\rm tot}&=&{-(1+\kappa)\mu^2+2\kappa\mu(1-2\Omega^2)+4\kappa\Omega^2\over
  1-\kappa}, \\
\tilde J_{\rm tot}&=&{4\kappa(1-\mu)\Omega\sqrt{\mu+\Omega^2}\over 1-\kappa}.
\end{eqnarray}

The loci of $\tilde
J_{\rm tot}=\pm M_{\rm tot}$ are the two intersecting critical curves
$\Omega=\pm\Omega_c(\mu)$ with
\begin{equation}
\Omega_c(\mu)={2\kappa-(1+\kappa)\mu \over
  \sqrt{8\kappa(1-\kappa)(1-\mu)}}.
\end{equation}
They cross at
\begin{equation}
\mu_c={2\kappa\over 1+\kappa},
\end{equation}
which is inside the strip for all $0<\kappa<1$, and they intersect the
edge $\mu=0$ of the strip at
\begin{equation}
\Omega_0=\sqrt{\kappa\over 2(1-\kappa)}.
\end{equation}
Hence for all physical values of $\kappa$ the strip contains regions
corresponding to point-particle, black hole and overspinning values of
the pair $(\tilde J,M)$, as we have already shown in general. 

\begin{figure}
\includegraphics[scale=0.5]{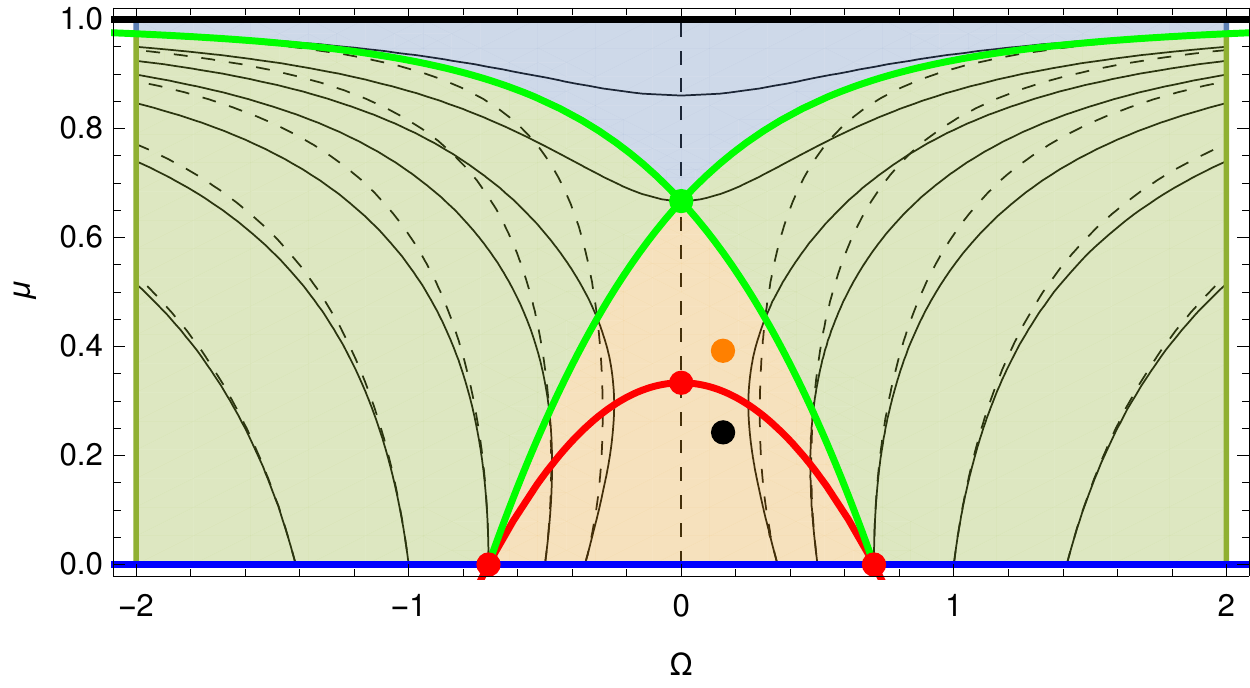}
\caption{The nature of the asymptotic metric for star-like solutions
  with the linear equation of state $p=\rho/2$. All solutions lie in
  the strip $0<\mu<1$, $-\infty<\Omega<\infty$. The asymptotic metric
  is of black hole type in the orange (bottom) region, of
  point-particle type in the blue (top) region and of overspinning
  type in the green (left and right) regions. The parameter values of
  the two solutions shown in Fig.~\ref{fig:fig4}, and which have the
  same $(\tilde J,M)$, are indicated by an orange and a black dot. The
  contours of $M=-1,-1/2,0,1/2,1,2,4,8,16$ (from top to bottom, solid)
  and $\tilde J=0,\pm 1/2,\pm 1,\pm 2, \pm 4,\pm 8,\pm 16$ (outward
  from the centre, dashed) are also shown. The crossing green lines
  indicate $\tilde J=\pm M$. The bottom region is split into two
  regions by the red line, each of which covers the same region in the
  $(\tilde J,M)$ plane. Solutions in the bottom half, such as the one
  indicated by the black dot, are conjectured to be unstable. The
  green dot is at $(0,\mu_c)$, and the three red dots are at
  $(\pm\Omega_0,0)$ and $(0,\mu_r)$.}
\label{fig:fig1}
\end{figure}

The parameter space $0<\mu<1$, $-\infty<\Omega<\infty$ of solutions is
shown in Fig.~\ref{fig:fig1} for $\kappa=1/2$, together with contour
lines of $M$ and $\tilde J$, the lines $|\tilde J|=|M|$, colour-coding
of the asymptotic metric as black-hole, point particle or
overspinning, and the curve that divides the black-hole region of
parameter space into two halves that cover the corresponding region of
$(\tilde J,M)$ space twice. This second curve is given by
\begin{equation}
\Omega_r^2(\mu)={1+\kappa\over 2(1-\kappa)}(\mu_r-\mu), \quad
\mu_r:={\kappa\over 1+\kappa}
\end{equation}
for $0<\mu<\mu_r$. We can deparameterise this curve to obtain $J^2$ as a
function of $M$ involving only square roots, but the result is messy. 

Solutions of black-hole type exist only for
$M<M_0$ with
\begin{equation}
M_0=8\Omega_0^4={2\kappa^2\over(1-\kappa)^2}.
\end{equation}
The maximum possible mass without
rotation is 
\begin{equation}
M_r=A(\mu_r)\mu_r={\kappa^2\over 1-\kappa^2}.
\end{equation}

%%%%%%%%%%%%%%%%%%%%%%%%%%%%%%%%%%%%%%%%%%%%%%%%%%%%%%%%%%%%%%%%%%%%%%
%%%%%%%%%%%%%%%%%%%%%%%%%%%%%%%%%%%%%%%%%%%%%%%%%%%%%%%%%%%%%%%%%%%%%%

\begin{figure}
\includegraphics[scale=0.5]{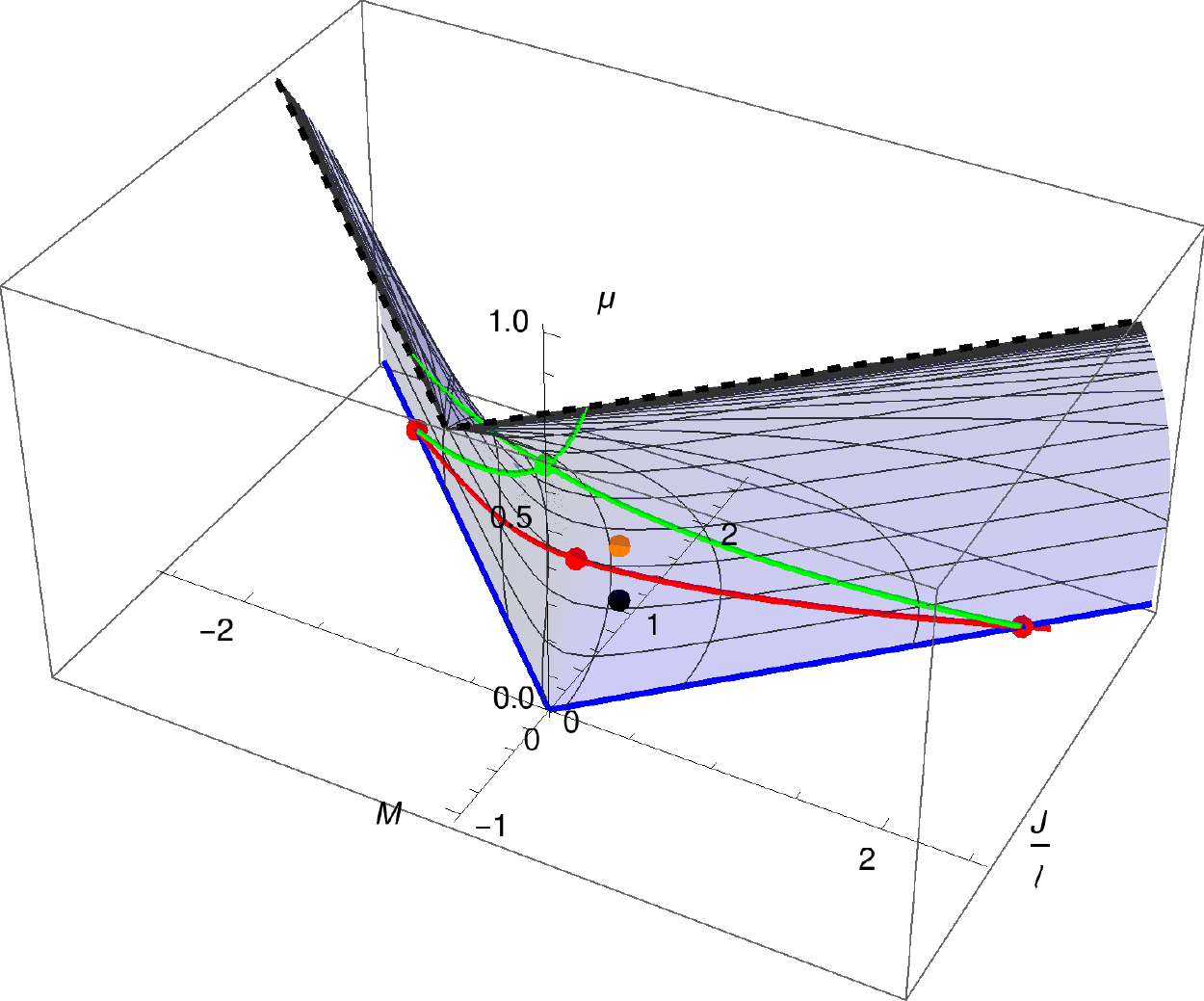}
\caption{Parametric plot of $(\tilde J,M)$ as a function of
  $(\mu,\Omega)$, embedded in three dimensions as $(\tilde
  J,M,\mu)$. All dots and thick curves correspond to those of the same
  colour in Fig.~\ref{fig:fig1}. Contours of $\Omega$ and $\mu$ are
  shown as thin lines. The thick red line denotes the locus of
  $|\partial(\tilde J,M)/\partial(\mu,\Omega)|=0$, where the embedded
  surface is vertical. The intersecting thick green lines denote the
  loci of $\tilde J=\pm M$ at nontrivial values of $\mu$. The bottom
  edge of the plot, $\mu=0$, is at $\tilde J=\pm M$, for $M>0$. The
  top edge of the plot (dashed black line), $\mu=1$ is at $\tilde
  J=\pm (M+1)$, for $M\ge -1$, with $M=-1$ only at $\mu=1$. The single
  point $M=-1$, $\tilde J=0$ in this plot corresponds to a 2-parameter
  family of test fluid solutions. Solutions in the area below the red
  line are conjectured to be unstable. The orange dot and the black
  dot represent two solutions with the same $M$ and $\tilde J$ that
  are presumed stable and unstable, respectively. The green dot is at
  $(0,M_c,\mu_c)$, and the three red dots are at $(\pm M_0,M_0,0)$ and
  $(0,0,\mu_r)$.}
\label{fig:fig2}
\end{figure}

\begin{figure}
\includegraphics[scale=0.5]{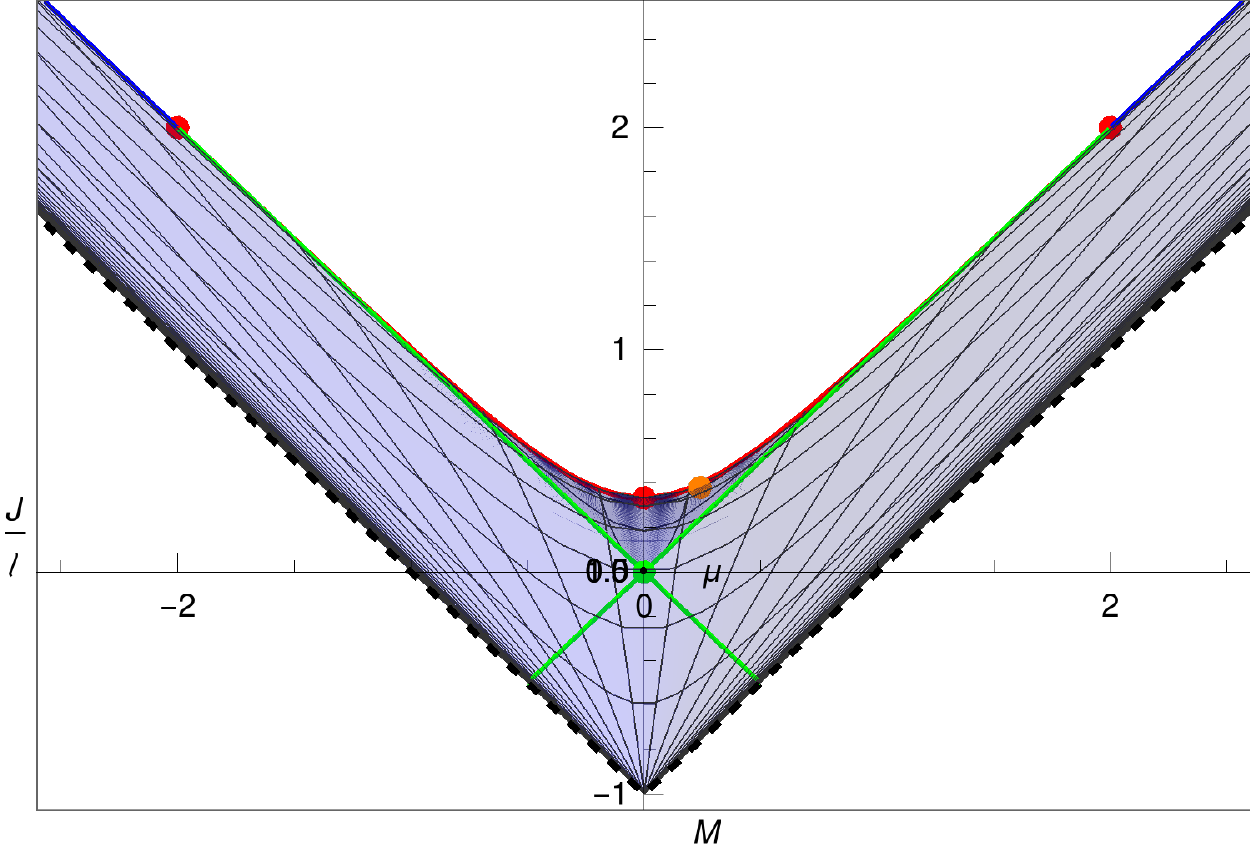}
\caption{A top view, suppressing the dimension $\mu$, of the plot in
  Fig.~\ref{fig:fig2}. All dots and curves are as described in
  Fig.~\ref{fig:fig2}. Note that the orange
  dot lies on top of, and so hides, the black one.}
\label{fig:fig3}
\end{figure}

The manifold of solution is shown embedded in $(\tilde
J,M,\mu)$ space in Fig.~\ref{fig:fig2} to show the double cover more
clearly, using the same colour-coding. A top view, suppressing the
$\mu$ direction and thus hiding the double cover, is given in
Fig.~\ref{fig:fig3}. 

%%%%%%%%%%%%%%%%%%%%%%%%%%%%%%%%%%%%%%%%%%%%%%%%%%%%%%%%%%%%%%%%%%%%%%
%%%%%%%%%%%%%%%%%%%%%%%%%%%%%%%%%%%%%%%%%%%%%%%%%%%%%%%%%%%%%%%%%%%%%%

%%%%%%%%%%%%%%%%%%%%%%%%%%%%%%%%%%%%%%%%%%%%%%%%%%%%%%%%%%%%%%%%%%%%%%
%%%%%%%%%%%%%%%%%%%%%%%%%%%%%%%%%%%%%%%%%%%%%%%%%%%%%%%%%%%%%%%%%%%%%%

\begin{figure}
\includegraphics[scale=0.5]{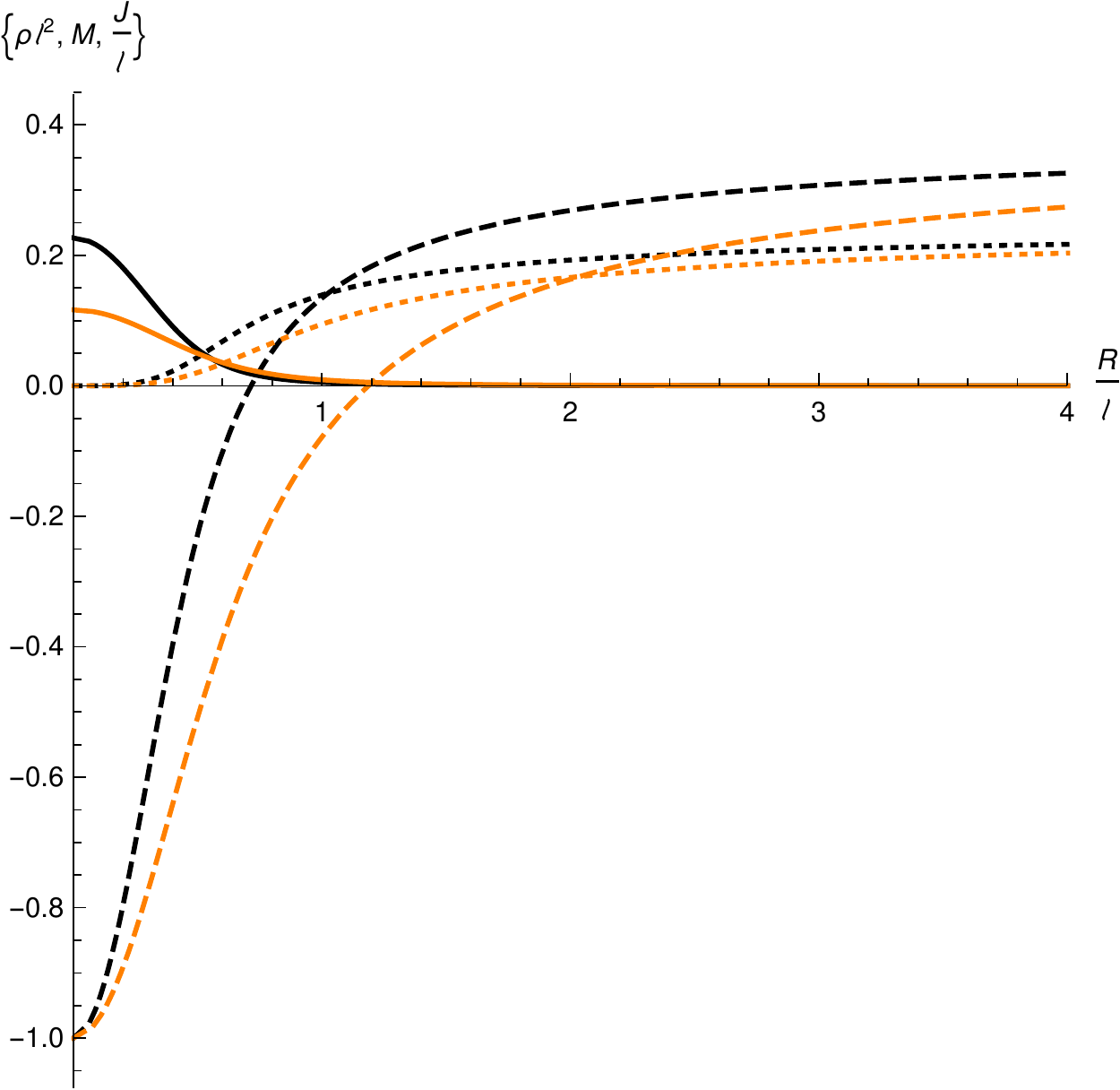}
\caption{An example of two star-like solutions with the equation of
  state $p=\rho/2$. Both have $M=0.38$ and $\tilde J=0.24$, but
  different central densities. We plot $\ell^2\rho$ (solid), $M$
  (dashed) and $\tilde J$ (dotted) against $R/\ell$.  The less compact
  solution, with $(\Omega,\mu)\simeq (0.154,0.242)$ and lower central
  density, (stable in nonlinear numerical time evolutions) is plotted
  in orange, and the more compact one with
  $(\Omega,\mu)\simeq(0.153,0.392)$ and higher density, (numerically
  found to be unstable) in blue.}
\label{fig:fig4}
\end{figure}

In all these figures, we have marked a specific pair of solutions with
black-hole class asymptotic metrics, both of which have the same total
mass $M=0.38$ and angular momentum $\tilde J=0.24$, but which have
different parameter values $(\Omega,\mu)\simeq (0.154,0.242)$ and
$(0.153,0.392)$. These solutions themselves are illustrated in
Fig.~\ref{fig:fig4} by plotting $M$, $\tilde J$ and $\ell^2\rho$ as
functions of $R/\ell$.

%%%%%%%%%%%%%%%%%%%%%%%%%%%%%%%%%%%%%%%%%%%%%%%%%%%%%%%%%%%%%%%%%%%%%%
%%%%%%%%%%%%%%%%%%%%%%%%%%%%%%%%%%%%%%%%%%%%%%%%%%%%%%%%%%%%%%%%%%%%%%

%%%%%%%%%%%%%%%%%%%%%%%%%%%%%%%%%%%%%%%%%%%%%%%%%%%%%%%%%%%%%%%%%%%%%%%%%%

\subsection{Modified linear equation of state $p=\kappa(\rho-\rho_s)$}

%%%%%%%%%%%%%%%%%%%%%%%%%%%%%%%%%%%%%%%%%%%%%%%%%%%%%%%%%%%%%%%%%%%%%%%%%%

A simple equation of state that admits solutions with a surface at
finite radius is the inhomogeneous linear one, 
\begin{equation}
p=\kappa(\rho-\rho_s),
\end{equation}
for $0<\kappa<1$ and $\rho_s\ge 0$.  Obviously this reduces to the
previous example for $\rho_s=0$. Proceeding as before, we find
\begin{equation}
\rho=\rho_0 \,y^{-{1+\kappa\over \kappa}}+{\kappa\rho_s\over 1+\kappa}
\left(1-y^{-{1+\kappa\over \kappa}}\right).
\end{equation}
We then obtain
\begin{equation}
\label{x2ymodlineos}
x^2=\tilde\mu(y^2-1)+{2\kappa(1-\tilde\mu)\over
  1-\kappa}\left(1-y^{-{1-\kappa\over\kappa}}\right)
\end{equation}
which is just (\ref{x2ylineos}) again, only with $\mu$ replaced by
\begin{equation}
\tilde\mu:=\mu-\sigma, 
\end{equation}
where 
\begin{equation}
\quad \sigma:={\kappa\over 1+\kappa} 8\pi s^2\rho_s
={\kappa\over 1+\kappa} (\mu+\Omega^2)8\pi \ell^2\rho_s.
\end{equation}
The other metric components follow, and we do not give them here. The
stellar surface is now at finite radius
\begin{equation}
y_*(\rho_0)=\left({(1+\kappa)\rho_0\over\rho_s}-\kappa\right)^{\kappa\over
  1+\kappa}.
\end{equation}
Note that $y_*(\rho_s)=1$ as expected. We have
\begin{equation}
m=2{1-\mu\over 1-\kappa}+{1+\kappa\over 1-\kappa}\sigma 
\left(1-\left({1-\mu+\sigma\over\sigma}\right)^{2\kappa\over
  1+\kappa}\right),
\end{equation}
which now depends also on $\Omega$ through $\sigma(\mu,\Omega)$. We
do not write down further expressions, which are complicated and do
not add new insight.

%%%%%%%%%%%%%%%%%%%%%%%%%%%%%%%%%%%%%%%%%%%%%%%%%%%%%%%%%%%%%%%%%%%%%%%%%%

\subsection{Polytropic equation of state $p=K\rho^k$}

%%%%%%%%%%%%%%%%%%%%%%%%%%%%%%%%%%%%%%%%%%%%%%%%%%%%%%%%%%%%%%%%%%%%%%%%%%

For
\begin{equation}
p=K\rho^k,
\end{equation}
the star has a surface at finite radius
\begin{equation}
y_*(\rho_0)=\left(1+K\rho_0^{k-1}\right)^{{k\over k-1}}
\end{equation}
if and only if $k>1$, consistent with the analysis in
Sec.~\ref{section:stars}. We find 
\begin{equation}
\rho(\rho_0;y)=K^{-{1\over k-1}}
\left(\left({y\over y_*(\rho_0)}\right)^{-{k-1\over k}}-1\right)^{1\over k-1},
\end{equation}
The functions $f(y)$ and hence $x^2(y)$ can be expressed in closed
form in terms of hypergeometric functions, as already noticed in
\cite{Cataldo}. The same is true for $m$, and hence $M_{\rm tot}$ and
$J_{\rm tot}$. We do not write down these expressions as they do not
give further insight.

%%%%%%%%%%%%%%%%%%%%%%%%%%%%%%%%%%%%%%%%%%%%%%%%%%%%%%%%%%%%%%%%%%%%%%%%%%

\section{Conclusions}

%%%%%%%%%%%%%%%%%%%%%%%%%%%%%%%%%%%%%%%%%%%%%%%%%%%%%%%%%%%%%%%%%%%%%%%%%%

We have constructed rotating perfect fluid star-like solutions in
2+1-dimensional general relativity with a negative cosmological
constant $\Lambda<0$. We defined these to have a regular centre, and
finite mass $M$ and spin $J$ at infinity. (We again suppress the
suffix ``tot'' in this Section.) We have found these solutions in
standard polar-radial coordinates $(t,R,\theta)$, in terms of two free
parameters $\mu$ and $\Omega$ that control their mass and spin, and we
have given expressions for the total mass $M$ and spin $J$ in terms of
the two free parameters.  We have thus established that star-like
solutions in 2+1 dimensions exist for generic equations of state.

Furthermore, we have shown that these solutions are analytic in
suitable coordinates, including at the centre, for analytic equations
of state (except at the surface, if there is a sharp surface). We have
also shown that their causal structure is that of the adS3 cylinder,
without closed timelike curves. 

For any equation of state with $0<p'(\rho)<1$ and where either
$p\sim\rho^k$ with $k>1$ as $\rho\to 0$, or $p=0$ occurs at finite
$\rho$, we have shown that rotating and non-rotating stars with a
sharp surface exist. The spacetime in the vacuum exterior is then the
BTZ solution. In the limiting case where the equation of state is
linear at low density, $p\simeq \kappa\rho$ with $0<\kappa<1$ as
$\rho\to 0$, the density goes to zero only asymptotically, but
sufficiently fast so that the spacetime is asymptotically BTZ with
finite $M$ and $J$.

We stress that the necessary and sufficient criterion for the existence
of stars with a surface at finite radius and finite $M$ and $J$ is
simply that the integral (\ref{ystareos}) converges at $p=0$. We have
not assumed further constraints on the equation of state except the
causality constraint $0<p'(\rho)<1$ for all $p>0$. 

We have shown that for a generic equation of state the $(\Omega,\mu)$
parameter space contains exterior/asymptotic metrics of all three BTZ
types: black-hole, point-particle and overspinning, but not for all
values $(\tilde J,M)$. More precisely, solutions for generic equations
of state cover all of the infinite region (\ref{chevron}) of the $(\tilde
J,M)$ plane, and a finite region bounded by
  (\ref{tilderedcurve}). In this second region, there are two
solutions for the same values of $M$ and $\tilde J$, with the more
compact one conjectured to be unstable.

For an arbitrary barotropic equation of state $p=p(\rho)$ our
solutions are in implicit form, involving two integrals and one
function inversion. The integrals can be solved in closed form for the
linear equation of state $p=\kappa\rho$, explicitly constructing the
space of solutions, and we have shown that this is possible also for
two other simple equations of state in which stars have sharp surfaces.

In spite of the local triviality of gravity, two compact
self-gravitating objects in 2+1 dimensions can interact
gravitationally through global effects \cite{DeserJackiwTHooft84} and,
for $\Lambda<0$, even merge to form a black hole; see \cite{HM99} for
an explicit construction of a spacetime representing the formation of
a spinning black hole from two massless point particles colliding with
impact parameter. However, because there are no tidal forces or
gravitational waves, unless and until the two objects actually touch 
they do not affect each other's local dynamics. In particular, if they
start in an axistationary state they remain so unless and until they
touch. This makes axistationary matter solutions even more relevant
for representing interacting compact objects than they are in 3+1
dimensions.

\acknowledgments

The early stage of this work was partly funded by a 2019 EPSRC
Vacation Bursary to the Unversity of Southampton. CG gratefully
acknowledges conversations with Jorma Louko about 2+1-dimensional
gravity. 

%%%%%%%%%%%%%%%%%%%%%%%%%%%%%%%%%%%%%%%%%%%%%%%%%%%%%%%%%%%%%

%%%%%%%%%%%%%%%%%%%%%%%%%%%%%%%%%%%%%%%%%%%%%%%%%%%%%%%%%%%%%%%%%%%%%%
\end{document}